\documentclass[conference]{IEEEtran}
\IEEEoverridecommandlockouts
\usepackage{cite}
\usepackage{amsmath,amssymb,amsfonts,bm}
\usepackage{algorithmic}
\usepackage{graphicx}
\usepackage{float}
\usepackage{subfigure}
\usepackage{textcomp}
\usepackage{xcolor}
\usepackage{enumitem}

\usepackage[linesnumbered,ruled,vlined]{algorithm2e}

\def\BibTeX{{\rm B\kern-.05em{\sc i\kern-.025em b}\kern-.08em
		T\kern-.1667em\lower.7ex\hbox{E}\kern-.125emX}}
\begin{document}
	
	\title{UAV-assisted Emergency Integrated Sensing and Communication Networks: A CNN-based Rapid Deployment Approach\\}
	
	\author{
		\IEEEauthorblockN{Zao Wang\IEEEauthorrefmark{1}, Lianming Xu\IEEEauthorrefmark{1}, Luyang Hou\IEEEauthorrefmark{1}, Ruoguang Li\IEEEauthorrefmark{2}, and Li Wang\IEEEauthorrefmark{1}\\}
		\IEEEauthorblockN{\IEEEauthorrefmark{1}School of Computer Science, Beijing University of Posts and Telecommunications, Beijing, China\\}
		\IEEEauthorblockN{\IEEEauthorrefmark{2}National Mobile Communications Research Laboratory, Southeast University, Nanjing, China\\}
		\IEEEauthorblockN{\{wangzao, xulianming, luyang.hou, liwang\}@bupt.edu.cn, ruoguangli@seu.edu.cn\\
		}
		\thanks{This work was supported in part by the National Key Research and
			Development Program of China under Grant 2020YFC1511801, and in part
			by National Natural Science Foundation of China under grants 62171054,
			62201071, 62201089 and U2066201. (Corresponding author: Li Wang.)}
	}
	
	\maketitle
	
	\begin{abstract}
	UAV-assisted integrated sensing and communication (ISAC) network is crucial for post-disaster emergency rescue. The speed of UAV deployment will directly impact rescue results. However, the ISAC UAV deployment in emergency scenarios is difficult to solve, which contradicts the rapid deployment. In this paper, we propose a two-stage deployment framework to achieve rapid ISAC UAV deployment in emergency scenarios, which consists of an offline stage and an online stage. Specifically, in the offline stage, we first formulate the ISAC UAV deployment problem and define the ISAC utility as the objective function, which integrates communication rate and localization accuracy. Secondly, we develop a dynamic particle swarm optimization (DPSO) algorithm to construct an optimized UAV deployment dataset. Finally, we train a convolutional neural network (CNN) model with this dataset, which replaces the time-consuming DPSO algorithm. In the online stage, the trained CNN model can be used to make quick decisions for the ISAC UAV deployment. The simulation results indicate that the trained CNN model achieves superior ISAC performance compared to the classic particle swarm optimization algorithm. Additionally, it significantly reduces the deployment time by more than 96\%.
	
	\end{abstract}
	
	\begin{IEEEkeywords}
		Emergency rescue, ISAC, UAV deployment, CNN.
	\end{IEEEkeywords}
	
	\section{Introduction}
	After natural disasters occur, the existing network infrastructure is severely damaged \cite{d2}. Building an emergency sensing and communication (ISAC) network is helpful for the smooth implementation of subsequent rescue operations, where the emergency ISAC network is an emergency task-oriented wireless communication network with sensing capabilities. Recently, unmanned aerial vehicles (UAV) have been widely used in the construction of emergency ISAC networks due to their high mobility and flexibility \cite{d1}. When building a UAV-assisted emergency network, there are two key issues to consider:  deployment and resource allocation.
	
	The UAV deployment and resource allocation has been investigated from several perspectives. Ref. \cite{a3} employed gibbs sampling and block coordinate descent to jointly optimize UAV locations, bandwidth, and power allocation to maximize network throughput while ensuring localization accuracy constraints are met. Ref. \cite{a4} proposed an approach based on genetic algorithms and multi-intelligence deep reinforcement learning to optimize UAV locations in order to maximize the ISAC performance. Ref. \cite{a41} develop a method based on the bipartite graph to solve the problem of optimal matching for a wireless communication network. Ref. \cite{a42} proposed a roth and vande vate-based distributed scheme to solve the dynamic matching problem in device-to-device communications network. The above-mentioned method requires a lot of time to compute UAV locations and resource allocation results. Whenever the input information of algorithms is changed, it needs to restart the computation. However, emergency scenarios are characterized by complex environments and dynamic user locations, which require the execution time of algorithms to be as short as possible. Therefore, the above-mentioned method is difficult to apply in emergency scenarios.
	
	Several studies have also investigated rapid UAV deployment. Ref. \cite{a5} adopted KMeans to solve the UAV communication deployment problem. Ref. \cite{a6} developed a method based on virtual force field to optimize UAV locations with the aim of maximizing the network throughput. However, these methods did not yield satisfactory network performance. Ref. \cite{a7} proposed a UAV deployment algorithm based on deep neural networks (DNN). This algorithm builds the optimal UAV locations dataset through reinforcement learning, and then trains a DNN model to compute UAV locations. However, DNN cannot well understand the two-dimensional user distribution information. In \cite{a8}, a learning framework based on CNN is designed to address the cache placement and trajectory optimization problem with the aim of maximizing the overall network throughput in vehicular networks. CNN can understand user distribution information, but CNN is primarily sensitive to image data. However, in emergency scenarios where personnel distribution is sparse, traditional CNNs have limitations. In addition, there is currently no rapid UAV deployment algorithm for ISAC network.
	
	Based on aforementioned analysis, this paper solves the rapid ISAC UAV deployment problem by proposing a CNN-based algorithm. The main contributions of this paper are summarized as follows:
	\begin{itemize}
		\item To solve the above problem in real time, we propose a two-stage framework based on CNN, which consists of an offline and an online stage. In the offline stage, we construct a UAV optimized deployment dataset for complex ISAC network scenarios and train a CNN deployment model based on this dataset. In the online stage, we utilize the trained CNN model to perform fast UAV deployment decisions.
		\item In the offline stage, we design a DPSO algorithm to compute UAV locations. Specifically, we introduce a nonlinear weight and learning factor update strategy to speed up the convergence and prevents algorithm falling into local optima. 
		\item To solve the problem caused by sparse distribution of users in emergency scenario, we further propose a feature enhancement method based on gaussian distribution to make CNN understand user distribution better.
	\end{itemize}
	
	\section{System Model and problem formulation}
	As illustrated in Fig. 1, we consider a multi-UAV-assisted ISAC system for emergency rescue in a mountainous forest area, where sensing only considers localization. Specifically, the system consists of a control vehicle and $M$ UAVs, in which the control vehicle serves as a centralized controller to schedule all the UAVs, and UAVs provide simultaneous communication and localization service for $N$ on-ground affected users. In this system, a signal service cycle is divided into two phases in time order: the communication phase and the localization phase. During the communication phase, each user can only establish a communication connection with one UAV. During the localization phase, Each user establishes localization connections with three UAVs. All UAVs share the same frequency band. Each UAV BS provides communication and localization services for users by frequency division multiple access technology (FDMA). The UAV and user sets are denoted by $\mathcal{M}=\left\{ 1,2,\cdots ,M \right\}$ and $\mathcal{N}=\left\{ 1,2,\cdots ,N \right\}$, respectively. The coordinates of $m$-th UAV and ground $n$-th user can be  represented as ${\mathbf{u}}_{m}=[{x_{m}},{y_{m}},{z_{m}}]$ and ${\mathbf{o}}_{n}=[{x_{n}},{y_{n}},0]$, respectively. The distance between $m$-th UAV and $n$-th user can be expressed as $ {{d}_{m,n}}={\left \| \mathbf{u}_{m}-\mathbf{o}_{n} \right \|}_2$.

	In practice, the location of UAVs is also subject to collision avoidance constraints and safety altitude constraints, i.e.,
	\begin{equation}
		{{d}_{m_1,m_2}}\ge {{d}_{min}},\forall m_1\in \mathcal{M},\forall m_2\in \mathcal{M},m_1\ne m_2\text{,}
	\end{equation}
	\begin{equation}
		{{z}_{m}}\ge {{h}_{\min }},\forall m\in \mathcal{M}\text{,}
	\end{equation}
	where $d_{min}$ denotes the minimum inter-UAV distance to ensure collision avoidance, $h_{min}$ is the minimum flight altitude to prevent collision with the mountain.
	
	The total path loss (in dB) between $m$-th UAV and $n$-th user in forest environments can be represented as follows, according to \cite{b1},
	\begin{small}
		\begin{equation}
			PL_{m,n}^{{}}=20\lg \left( \frac{4\pi f{{d}_{0}}}{c} \right)+10\eta \lg \left( \frac{{{d}_{m,n}}}{{{d}_{0}}} \right)+{{X}_{\sigma }}+PL_{m,n}^{Slant}\text{,}
		\end{equation}
	\end{small}where $f$ represents the carrier frequency, ${{d}_{0}}$ is the reference distance, and $c$ represents the speed of signal. $\eta$ is the path loss factor, $X_{\sigma}$ is the shadowing fading, which is a Gaussian random variable with  zero-mean and standard deviation $\sigma$. $PL_{m,n}^{Slant}$ represents the additional path loss in forest \cite{b3}.
	
	Therefore, the channel power gain between $m$-th UAV and $n$-th user can be expressed as
	\begin{equation}
		{{g}_{m,n}}={{10}^{\frac{-{PL}_{m,n}}{10}}}\text{.}
	\end{equation}
	\subsection{Communication Model}
	During the communication phase, we define the matrix ${{\mathbf{A}}}$ to represent the communication association between UAVs and users. Let $\alpha_{m,n}=1$ if $m$-th UAV provides communication service to $n$-th user, otherwise $\alpha_{m,n}=0$. When $\alpha_{m,n}=1$, we consider the signal-to-interference-plus-noise ratio (SINR) of $n$-th user receiving $m$-th UAV as
	\begin{equation}
		\gamma _{m,n}^{{}}=\frac{{{p}_{m,n}}{{g}_{m,n}}}{\sum\nolimits_{i\ne m}^{i\in \mathcal{M}}{{{p}_{i,n}}{{g}_{i,n}}+\psi}}\text{,}
	\end{equation}
	where ${{p}_{m,n}}$ is the transmitting power allocated by $m$-th UAV to $n$-th user, assuming that each UAV equally allocates the transmit power to the served users such that ${{p}_{m,n}}={}^{{{p}_{\max }}}/{}_{\sum\nolimits_{j\in \mathcal{N}}{\alpha_{m,j}}}$, where $p_{max}$ denotes the transmit power limit of the UAV. To ensure the communication signal quality for the served users, each UAV can service up to ${{K}_{C}}$ users in communication phase and the SINR for each user should be greater than ${{\gamma }_{C}}$. 
	
	We use throughput to measure the communication performance, the throughput of $n$-th user can be expressed as 
	\begin{equation}
		{{R}_{n}}={{b}_{m,n}}\log \left( 1+{{\gamma }_{m,n}} \right)\text{,}
	\end{equation}
	where ${{b}_{m,n}}$ is the bandwidth allocated by $m$-th UAV to $n$-th user, assuming that each UAV distributes its total bandwidth $b_{max}$ evenly across all the users it services such that ${{b}_{m,n}}={}^{b_{max}}/{}_{\sum\nolimits_{j\in \mathcal{N}}{\alpha_{m,j}}}$.
	
	\subsection{Localization Model}
	\begin{figure}[t]
		\centering
		\includegraphics[width=0.4\textwidth]{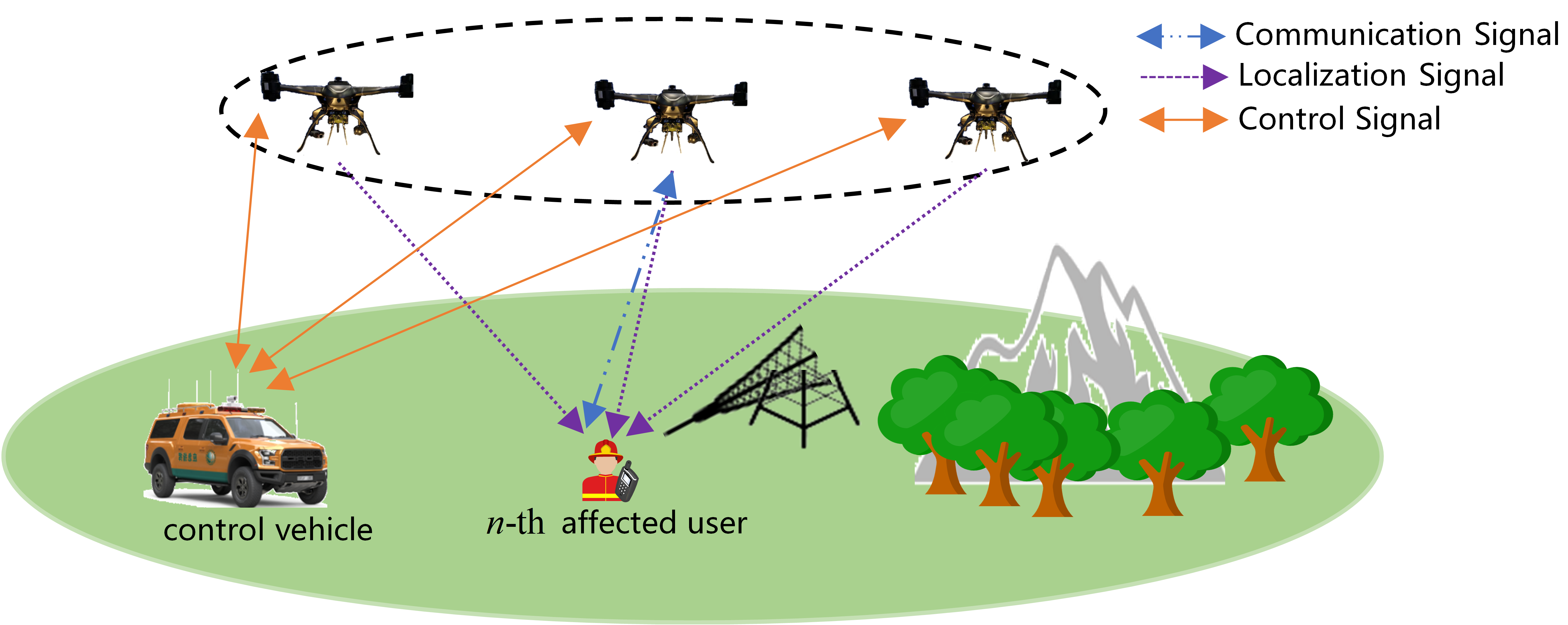}
		\caption{An overview of the UAV-assisted ISAC network}
	\end{figure}
	During the localization phase, the ground device calculates its location by estimating the time difference of arrival (TDOA) of signals emitted by the UAVs. Similarly, we define a matrix ${{\mathbf{B}}}$ to represent the location association between the UAVs and the users, where $\beta_{m,n}=1$ when $m$-th UAV provides location service for $n$-th user; otherwise $\beta_{m,n}=0$. To ensure the localization signal quality for the served users, each UAV is assumed to be able to serve up to ${K}_{P}$ users in the localization phase and the corresponding SINR must be greater than the threshold ${\gamma }_{P}$.
	
	 We adopt PDOP as the indicator to measure the localization accuracy. When the user's location can be calculated, we define the UAVs localization service group for $n$-th user as $\epsilon _{n}=\{{{m}_{i,n}}|i\in \{1,2,3\}\wedge {{m}_{i,n}}\in \mathcal{M}\}$, the PDOP of $n$-th user can be expressed as follows, according to \cite{b4},
	\begin{equation}
		{{\rho }_{n}}=\sqrt{tr\{[\mathbf{H}_n^T\mathbf{H}_n]^{-1}\}}\text{.}
	\end{equation}
	Assume that ${{m}_{n,1}}$-th UAV serves as the TDOA reference station in $\epsilon _{n}$, for simplicity of exposition, we define unit vectors $\mathbf{v}_{m,n} = \frac{\mathbf{u} _m-\mathbf{o} _n}{d_{m,n}} $, where $\mathbf{H}_n$ is given by
	\begin{equation}
	\mathbf{H}_{n} =\begin{bmatrix}
		\mathbf{v} _{m_{2,n},n}-\mathbf{v} _{m_{1,n},n} \\
		\mathbf{v} _{m_{3,n},n}-\mathbf{v} _{m_{1,n},n}
	\end{bmatrix}\text{.}
	\end{equation} 
	We define a piecewise function ${{G}_{n}}$ to represent the localization performance of $n$-th user
	\begin{equation}
		{{G}_{n}}=\left\{ \begin{matrix}
			{{\rho }_{n}}\text{,} & {{\rho }_{n}}\le {{\rho }_{\max }}\text{,} \\
			{{\rho }_{\max }}\text{,} & \text{otherwise}\text{,}\\
		\end{matrix} \right.
	\end{equation}
	where ${{\rho }_{\max }}$ is the upper limit threshold of PDOP.
	\subsection{Problem Formulation}
	We consider a weighted sum utility function to measure the ISAC performance, which is given by
	\begin{equation}
		{{\chi }_{n}}=(\lambda {{R}_{n}}+(1-\lambda )\frac{1}{{{G}_{n}}})\text{,}
	\end{equation}
	where $\lambda$ represents the weight of communication and localization performance. Because ${{R}_{n}}$ and $\frac{1}{{{G}_{n}}}$ have different dimensions and magnitudes, we use the max-min normalization method to standardize the two indicators.
	
	We hope to optimize UAV locations under a known UAV-user association strategy. Let $\mathcal{U}=\{{{\mathbf{u}}_{m}}|m\in \mathcal{M}\}$ represent the UAV locations, and $\mathcal{A}=\{{{\mathbf{A}}},{{\mathbf{B}}}\}$ represent the association strategy between UAVs and users. The goal is to jointly optimize $\mathcal{U}$ and $\mathcal{A}$, in order to maximize the ISAC performance of all users. This optimization problem can be formulated as
	\begin{subequations}
		\begin{align}
			& \underset{\begin{smallmatrix}\mathcal{U},\mathcal{A}\end{smallmatrix}}{\mathop{\max}}\,\sum\nolimits_{n\in \mathcal{N}}{{{\chi }_{n}}}  \\
			s.t.\quad & \sum\limits_{m=1}^{M}{\alpha_{m,n}}\le 1,\forall n\in \mathcal{N}\text{,}  \\ 
			& \sum\limits_{m=1}^{M}{\beta_{m,n}}\le 3,\forall n\in \mathcal{N}\text{,} \\
			& \sum\limits_{n=1}^{N}{\alpha_{m,n}}\le {{K}_{C}},\forall m\in \mathcal{M}\text{,} \\ 
			& \sum\limits_{n=1}^{N}{\beta_{m,n}}\le {{K}_{P}},\forall m\in \mathcal{M}\text{,}  \\
			& \gamma_{m,n}>{{\gamma }_{C}},\alpha_{m,n}=1\text{,} \\ 
			& \gamma_{m,n}>{{\gamma }_{P}},\beta_{m,n}=1\text{,} \\ 
			& {{z}_{m}}\ge {{h}_{\min }},\forall m\in \mathcal{M}\text{,} \\ 
			& {{d}_{m_1,m_2}}\ge {{d}_{min}},\forall m_1\in \mathcal{M},\forall m_2\in \mathcal{M},m_1\ne m_2\text{,}
		\end{align}
	\end{subequations}
	where (11b) indicates that each user can be served by at most one UAV, (11c) represents that at most three UAVs collaborate in locating each user, (11d) and (11e) represent the maximum number of communication and localization users that can be accessed by each UAV, respectively, (11f) and (11g) represent the minimum SINR for communication and localization signals of users, respectively, (11h) represents the minimum flight altitude, and (11i) represents the minimum safety distance between UAVs.
	
	\section{UAVs Deployment Based On CNN}
	The UAV deployment problem formulatd in Section II is a mixed integer nonlinear programming problem that is difficult to solve. Traditionally, solving this problem need to calculate ISAC performance of all users. However it will cost a lot of time. Therefore, We propose a new UAV deployment method named CNN-based Deployment Agent (CNNDA), which can achieve rapid UAV depolyment. Fig. 2 illustrates the processing flow of the proposed CNNDA, which consists of offline and online stage. In the offline stage, we design a DPSO algorithm to optimize the UAV locations. Based on this, we construct a dataset consisting of pairs of optimal UAV locations and user locations. Then, we train a CNN model using this dataset to learn the mapping function from users distribution to optimal UAV locations. In the online stage, we can quickly obtain UAV locations with the trained CNN model.
	
	\begin{figure*}[t]
		\centering
		\includegraphics[width=0.85\textwidth]{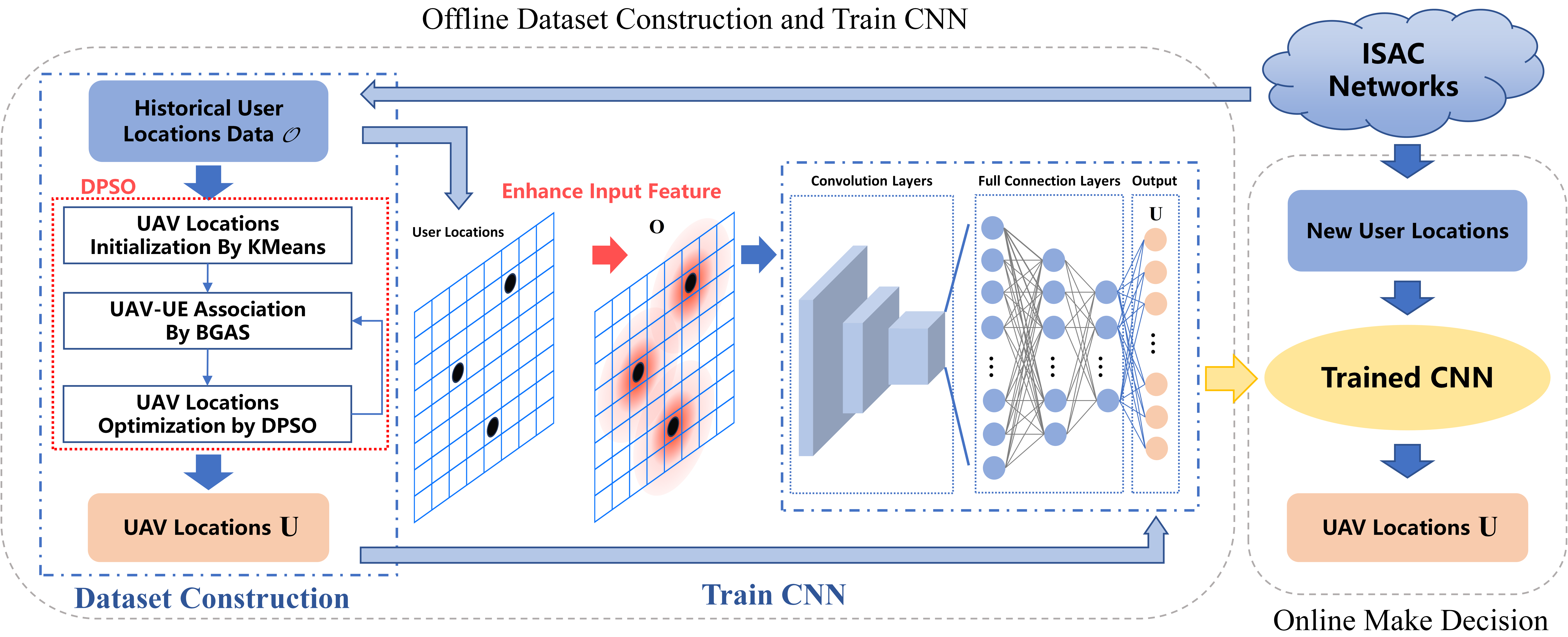}
		\caption{Process flow of the CNNDA framework}
	\end{figure*}
	
	\subsection{Dataset Construction}
	In the offline stage, we build an optimal UAV locations dataset to train CNN. Specifically, we first generate ${K_D}$ pieces of ground user locations according to random distribution, and we define the $k$-th data as ${\mathcal{O}_k} = \{\mathbf{o}_{k,1}, \cdots, \mathbf{o}_{k,N}\}$. Then, we compute ${K_D}$ corresponding optimal UAV locations according to the proposed DPSO. There are various methods to obtain the optimal UAV locations, such as exhaustive algorithm and heuristic algorithms. Considering exhaustive algorithm requires a lot of time and computational resources, we adopts a heuristic method to obtain the optimal UAV locations.
	
	The UAV deployment problem formulatd in Section II can be divided into two subproblems: one is to determine the association between UAVs and users $\mathcal{A}$, and another is to find the optimal UAV locaTtions $\mathcal{U}$. When solving $\mathcal{A}$, we need to ensure that $\mathcal{U}$ is fixed, and vice versa.
	
	\subsubsection{\textbf{UAV-UE Association}}
	To solve the subproblem of association between UAVs and users, We propose the benefit based greedy association strategy (BGAS) that takes into account both service quality and UAV capacity constraints. Specifically, we first define the reward functions of communication and localization for user $n$ as $q^C_n$ and $q^P_n$, respectively. Once $\mathcal{U}$ is determined, we calculate two benefits for all users within the communication or localization coverage range. Based on these benefits, we sort all users in descending order and associate the top-ranked user with corresponding UAV. 
	
	In the communication phase, we define the communication benefit $q_n^C$ for $n$-th user, which is expressed as
	\begin{equation}
		{q}_{n}^{C}={{\gamma }_{m,n}},\ m=\underset{i\in S_{n}}{\mathop{\arg \max }}\,\gamma _{i,n}^{{}}\text{,}
	\end{equation}
	\begin{equation}
		S_{n}=\left\{ i|i\in \mathcal{M}\wedge \gamma _{i,n}^{{}}>\gamma _{C}\wedge {{K}_{C}}(i)>0 \right\}\text{,}
	\end{equation}
	where ${{K}_{C}}(i)$ represents the remaining communication connections of $i$-th UAV, and $S_{n}$ represents the set of UAVs that can provide communication services for $n$-th user. 
	
	In the localization phase, we define the localization benefit of $n$-th user as $q_{n}^{P}$, which is represented as
	\begin{equation}
		q_{n}^{P}={{\rho }_{{{w}_{n}},n}},w_{n}^{{}}=\underset{w_{n}^{{}}\in {{W}_{\text{n}}}}{\mathop{\arg \min }}\,{{\rho }_{w_{n}^{{}},n}}\text{,}
	\end{equation}
	\begin{equation}
		{{W}_{n}}=\left\{ {\epsilon _{i}}|\forall m\in {\epsilon _{i}}\wedge {{\gamma }_{m,n}}>\gamma _{P}\wedge {{K}_{P}}(m)>0 \right\}\text{,}
	\end{equation}
	where ${{K}_{P}}(m)$ represents the remaining localization connection times for $m$-th UAV, and ${{W}_{n}}$ represents the set of UAVs that can provide localization services for $n$-th user.
	\subsubsection{\textbf{UAV deployment optimization}}
	In the subproblem of optimizing UAV locations, we propose a DPSO algorithm address issues such as local optima and slow convergence in traditional PSO algorithm. In DPSO, each particle location $l$ represents the location of all UAVs: $l=\left\{ {{\bm{u}}_{1}},{{\bm{u}}_{2}},\cdots {{\bm{u}}_{M}} \right\}$, the main process of  DPSO is as follows:
	
	2.a) \textbf{Initialization Stage}: The initial population contains $k$ particles, i.e., $\ell =\left\{ {{l}_{1}},{{l}_{2}},\cdots ,{{l}_{k}} \right\}$ . We use the KMeans algorithm to initialize the location of a certain particle $l_i$. Specifically, we calculate $M$ users cluster centers as the initial locations of the $M$ UAVs according to KMeans and set the value of $l_i$. Other particles are initialized randomly. Finally, we evaluate all particles and set each individual's optimal location $p_i$ and the population's optimal location $g$.
	
	2.b) \textbf{Evaluation Stage:} Based on BGAS, we can get the UAV-user association $\mathcal{A}$ for each particle. We then calculate the fitness value for each particle. We define a fitness function $F(l)$ to evaluate each particle, $F(l)$ is related to the optimization objective and its constraints, and it can be expressed as
	\begin{equation}
		F(l)=\left\{ \begin{matrix}
			(11a), & \text{(11h) and (11i),}  \\
			{{F}_{\min }}, & otherwise\text{,}  \\
		\end{matrix} \right.
	\end{equation}
	where $F_{min}$ is the minimum fitness in the current population.
	\begin{algorithm}[t]
		\caption{Dataset Construction in the Offline stage}
		\LinesNumbered
			Initialize \textbf{DPSO} parameters;
			
			\For{episode $k$:=$1$,...,$K_D$}{
				
				Generate user locations ${\mathcal{O}_k}$ randomly;
				
				Initialize particles by KMeans and Random;
				
				\For{epoch $t$:=$1$,...,$t_{max}$}{
					
					Calculate UAV-UE association $\mathcal{A}$ by \textbf{BGAS};
					
					Calculate $w(t)$ according to (19);
					
					Calculate $c_1(t)$, $c_2(t)$ according to (20);
					
					\For{each particle $i$}{
						
						Calculate fitness $F(l_i)$ according to (16);
						
						Update $v_i$, $l_i$ according to (17), (18) respectively;
						
						\If{$F(l_i) > F(p_i)$}{
							
							Update the individual best location $p_i$;
						}
						
						\If{$F(l_i) > F(g)$}{
							
							Update the population best location $g$;
						}
						
					}
				}

				Save the users’ location ${{\mathcal{O}}_{k}}$ and final UAVs’ location ${{\mathcal{U}}_{k}}=g$ as a piece of dataset;
			}
			
			Get a dataset $\mathbf{D}$.
	\end{algorithm}
	
	2.c) \textbf{Update Stage}: For particle $i$, if $F(l_i)$ is better than its individual best fitness $F(p_i)$, then update the individual best location $p_i$. If $F(l_i)$ is better than its population best fitness $F(g)$, then update the population best location $g$. The update strategy of velocity $v_i$ and location $l_i$ are expressed by
	\begin{small}
		\begin{equation} 
			v_{i}(t+1)=wv_{i}(t)+{{c}_{1}(t)}{{r}_{1}}({{p}_{i}}-l_{i}(t))+{{c}_{2}(t)}{{r}_{2}}({{g}_{i}}-l_{i}(t))\text{,}
		\end{equation}  
	\end{small}

	\begin{equation}
	l_{i}(t+1)=l_{i}(t)+v_{i}(t+1)\text{,}
	\end{equation}
	where $t$ represents the number of iterations, $w$ is the inertia weight, $c_1$ and $c_2$ are learning factors, $r_1$ and $r_2$ are randomly generated numbers between $[0, 1]$. Distinguishing from the traditional PSO algorithm, we introduce a non-linearly decreasing $w(t)$. In the early stages of the search, $w(t)$ is a larger value, allowing particles to explore the search space globally. In the later stages, $w(t)$ is a smaller value, facilitating local optimization, which is expressed as
	\begin{equation}
		{w(t)}=({{w}_{ini}}-{{w}_{end}}-0.2)\exp (\frac{1}{1+{7t}/{{{t}_{\max }}}})\text{,}
	\end{equation}
	where ${w}_{ini}$ represents the initial value of the inertia weight, ${w}_{end}$ represents the final value of the inertia weight, and $t_{max}$ represents the maximum number of iterations.
	
	For $c_1(t)$ and $c_2(t)$, when $c_1(t) \ge c_2(t)$, the particle tends to move to the individual's best location, conversely prefer population best. In DPSO, we make $c_1(t)$ non-linearly decreasing and $c_2(t)$ non-linearly increasing. This allows us to emphasize global search capability in the early stage and shift towards local search capability in the later stage. $c_1(t)$ and $c_1(t)$ share the same formula, which is expressed as
	\begin{equation}
		{{c}(t)}={{c}_{end}}+({{c}_{ini}}-{{c}_{end}})\frac{{{({{t}_{\max }}-t)}^{1.2}}}{{{t}^{1.2}}}\text{,}
	\end{equation}
	where ${c}_{ini}$ and ${c}_{end}$ respectively represent the initial value and final value of the learning factors during the iteration process.
	
	Algorithm 1 shows how to get a dataset. According to BGAS+DPSO, we can calculate the near-optimal UAVs deployment set ${\mathcal{U}_k}$ for each user information ${{\mathcal{O}}_k}$. By repeating BGAS+DPSO ${K_D}$ times, we can obtain the optimal strategy dataset $\mathbf{D}$ of ${K_D}$ samples, which can be represented as
	\begin{equation}
		\mathbf{D}=[({{\mathcal{O}}_{1}}, {{\mathcal{U}}_{1}}),\cdots ,(({{\mathcal{O}}_{{{K}_{D}}}},{{\mathcal{U}}_{{{K}_{D}}}}]\text{.}
	\end{equation}
	
	\subsection{Training the CNN}
	After obtaining the dataset $\mathbf{D}$, we train a CNN model to obtain a mapping function from users to UAV locations. Our model consist of input layer, feature layer and output layer.
	
	1) \textbf{Input}: We divide the ground area into $L\times L$ grids and define the $k$-th data's user distribution matrix $O_k$. In the input layer, we perform feature enhancement on the original user distribution data to solve gradient disappearance caused by the sparse users distribution. Specifically, for the grid position of $n$-th user $(i_n,j_n)$, and we use the gaussian enhancement method to calculate the pixel value of the $i$-th row and $j$-th column of the matrix, which can be expressed as
	\begin{equation}
		{{\mathbf{O}}_{k,(i,j)}}=\max \left\{ \exp (\frac{{{(i-{{i}_{n}})}^{2}}+{{(j-{{j}_{n}})}^{2}}}{-2{{\xi }^{2}}})|\forall n\in \mathcal{N} \right\}\text{,}
	\end{equation}
	where $\xi$ is the standard deviation of Gaussian enhancement. In this way, we enrich the image data without interfering with the original user distribution data, which is more conducive to the learning of CNN.
	
	\begin{figure}[t]
		\centering
		\includegraphics[width=0.35\textwidth]{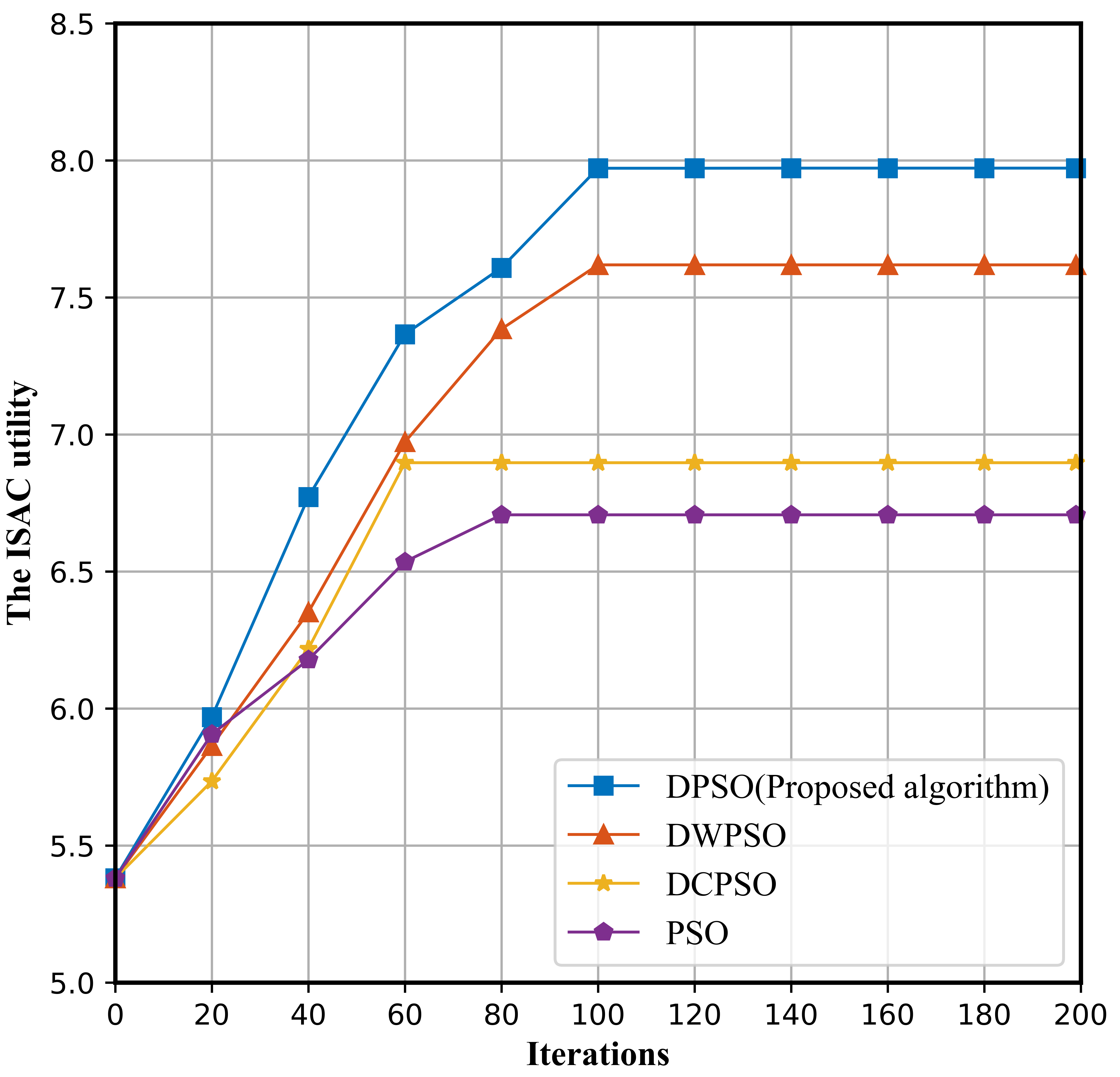}
		\caption{The convergence of the proposed algorithm and its benchmarks}
	\end{figure}
	
	2) \textbf{Network structure}: The CNN model consists of a feature extraction layer and a decision mapping layer. The feature extraction layer consists of two convolutional layers. The first convolutional layer consists of a 64-channel $7\times 7$ convolution kernel and a maximum pooling layer. The second The first convolutional layer consists of a 128-channel $3\times 3$ convolutional kernel and a maximum pooling layer. The decision mapping layer uses a multi-layer perceptron, and its input layer, hidden layer, and output layer are composed of 1024, 2048, and $3M$ neurons respectively; all neurons in the network use ReLU as the activation function.
	
	3) \textbf{Output}: The output of the CNN corresponds to all UAV locations, which can be represented as
	\begin{equation}
		\mathbf{U}=[{{x}_{1}},{{y}_{1}},{{z}_{1}};{{x}_{2}},{{y}_{2}},{{z}_{2}};\cdots ;{{x}_{M}},{{y}_{M}},{{z}_{M}}]\text{.}
	\end{equation}
	
	We adopt Mean Squared Error (MSE) as the loss function\cite{b6}. Additionally, we set pruning probabilities for each neuron to avoid overfitting problem\cite{b7}.
	\section{Simulation Results}
In this section, we evaluate the performance of the proposed method through simulations. We consider a $5$ km$\times$ $5$ km area. The default values of main parameters about UAVs are: $M=5$, $K_C=10$, $K_D=30$, $h_{min}=50$ m, $d_{min}=100$ m, $\gamma_C=3$ dB, $\gamma_P=1$ dB, $p_{max}=3$ w and $b_{max}=1$ MHz. For the forest channel, the default values of main parameters are: $A=0.25$, $C=0.39$, $E=0.25$, $G=0$, $H=0.05$, $\sigma=6$, $\eta=5$, $\psi=-140$ dbm and $f=1.4$ GHz. In the DPSO algorithm, the number of iteration is 200. We construct a dataset of 10,000 sample, and the ratio of training set to test set is 9:1. Since the CNNDA is a two-stage algorithm, we analyze and evaluate the algorithm performance from the offline and online stages.
	
	\begin{figure}[t]
		\centering
		\includegraphics[width=0.35\textwidth]{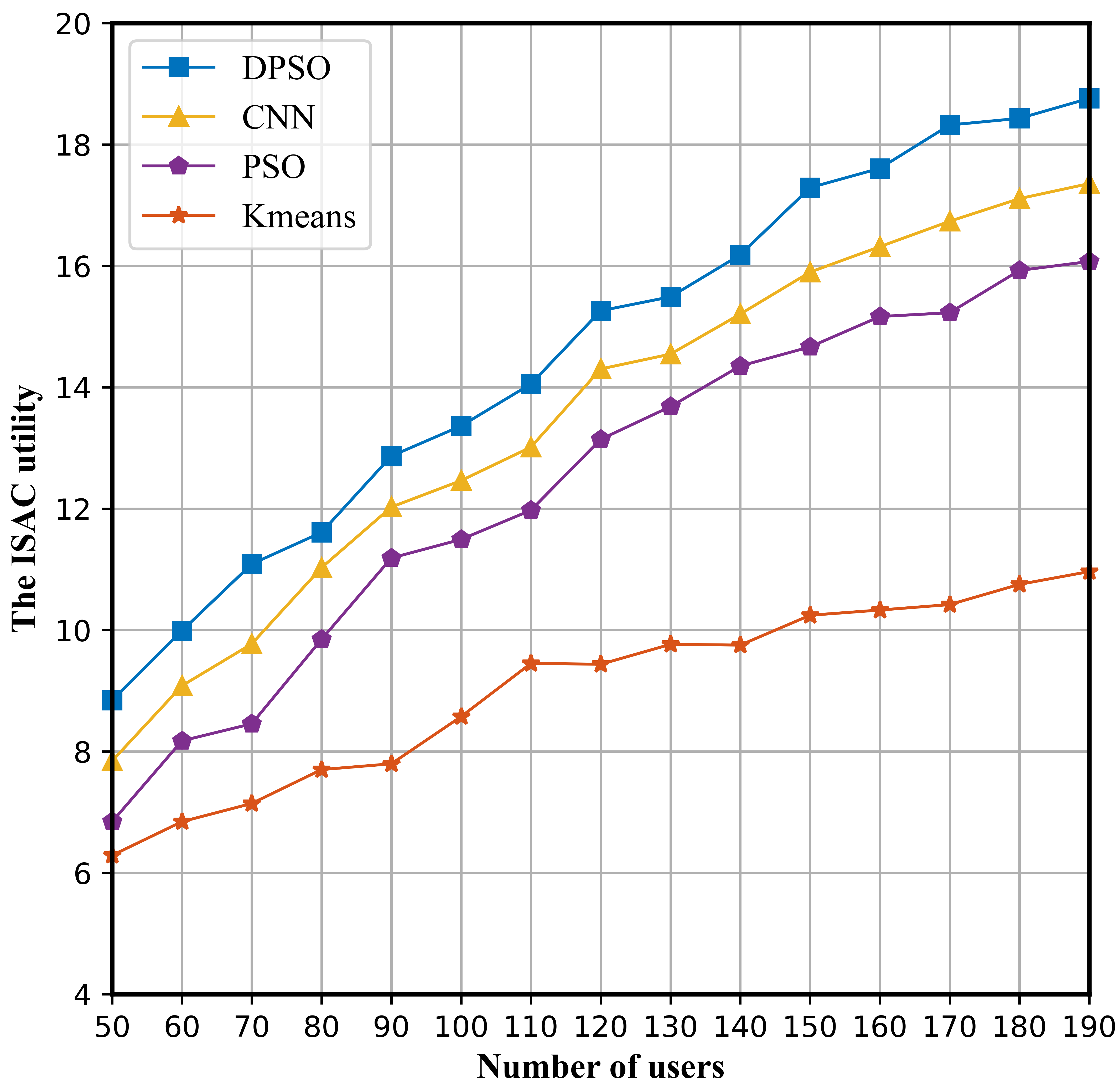}
		\caption{The ISAC performance versus the number of users.}
	\end{figure}
	
	For the offline stage, we compared three different algorithms, namely PSO, dynamic weight PSO (DWPSO), dynamic learning factor PSO (DCPSO) to verify the effectiveness of DPSO. Fig. 3 shows the convergence results of different PSO algorithms, and the vertical axis is the ISAC utility. After curve comparison, it can be observed that all algorithms converge before 100 generations, but compared with other three algorithm, DPSO can effectively help the algorithm jump out of the local optimal and improve the ISAC performance.
	
	For the online stage, we considered the generalization ability of the CNN model under different user quantity scenarios from the perspectives of ISAC performance and algorithm execution time. We selected the heuristic algorithm PSO and the machine learning algorithm Kmeans, as well as the proposed DPSO algorithm during offline optimization, as benchmark algorithms. In terms of ISAC performance, Fig. 4 shows our CNN model outperforms PSO and Kmeans in different user quantity scenarios, and its performance is closed to DPSO, which effectively learns the optimization ability of DPSO. In terms of algorithm execution time, Fig. 5 shows that our CNN model exhibits a deployment time close to the Kmeans, and its CPU execution time remains between 0.3-0.4s. Compared with heuristic algorithms such as DPSO and PSO, the CPU computing time of the proposed method is reduced by more than 96\%. Since the CNN model does not need to iteratively calculate the ISAC performance of each user, its deployment time is short and stable for any user scenario.
	
	\section{Conclusions}
	This paper investigates ISAC UAV deployment a in emergency scenarios with the aim of achieving desirable performance with short deployment time. The proposed CNNDA algorithm takes user locations as input and quickly determines UAV locations through the trained CNN model. This algorithm avoids computing the ISAC performance for each user, which is a time-consuming process. The simulation results show that the CNN model can achieve excellent ISAC performance on the one hand, and on the other hand, its execution time is maintained between 0.3-0.4s, which reduces the deployment time by more than 96\% compared with other iterative algorithms. Additionally, the CNN model exhibits excellent generalization ability, as its execution time remains stable and at a second-level across scenarios with varying numbers of users.
	
		\begin{figure}[t]
		\centering
		\includegraphics[width=0.35\textwidth]{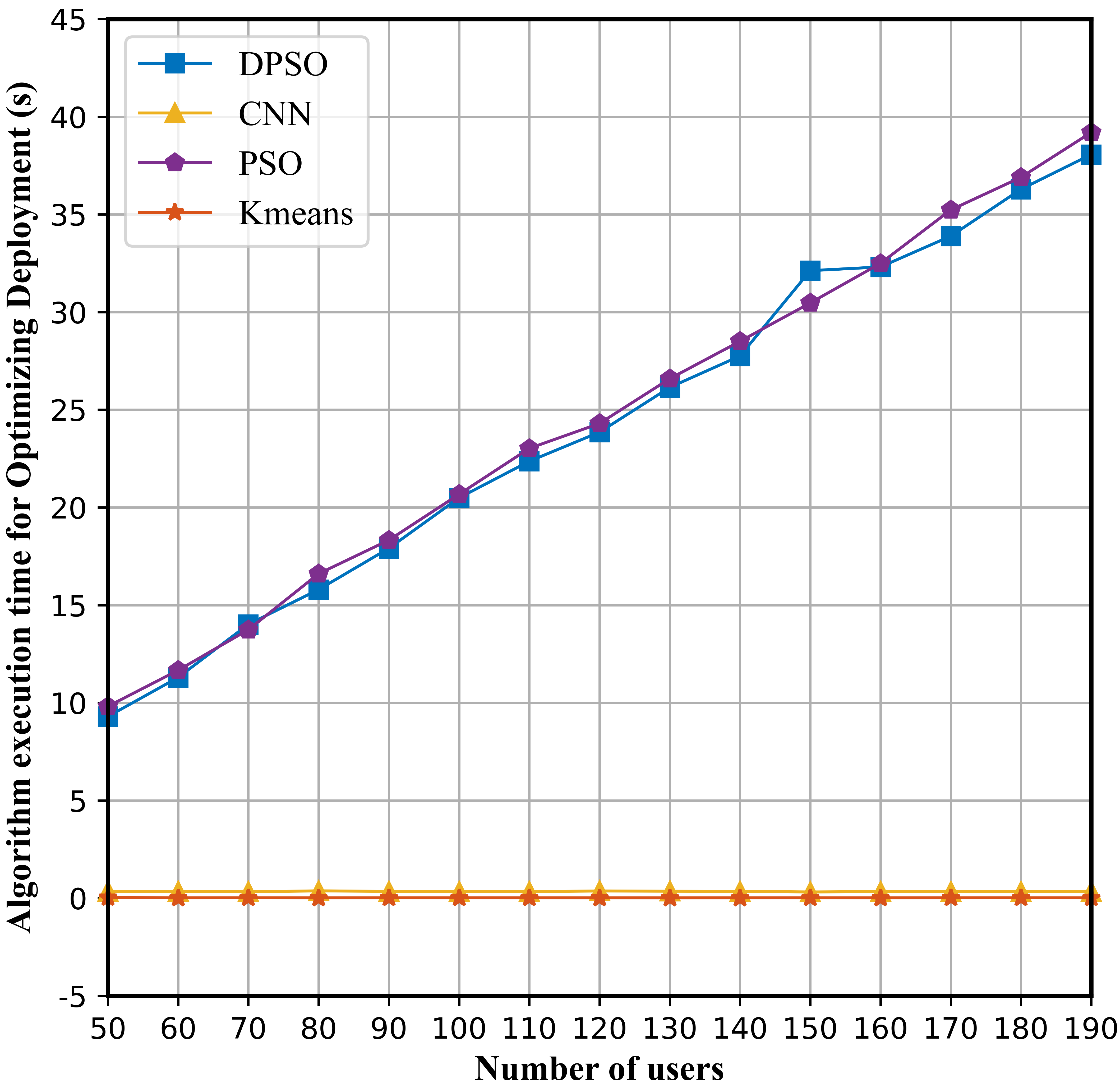}
		\caption{Algorithm execution time versus the number of users.}
	\end{figure}

\end{document}